\documentclass[UTF8,showpacs,superscriptaddress,reprint, amsmath,amssymb, aps,twocolumn]{revtex4}
\usepackage{amsmath}
\usepackage{graphicx}
\usepackage{dcolumn}
\usepackage{bm}
\usepackage{epsfig}
\usepackage{epsf}
\usepackage{color}
\usepackage{float}
\usepackage[colorlinks=true,
urlcolor=blue,
linkcolor=blue,
anchorcolor=blue,
citecolor=blue,
]{hyperref}
\usepackage{subfigure}
\allowdisplaybreaks[4]
\def \llb {\Lambda\bar\Lambda}

\def \ee {e^+e^-}

\def \ee {e^+e^-}

\usepackage[mathlines]{lineno}

\begin{document}

\preprint{APS/123-QED}

\title{Phenomenological study of $J/\psi\to\Xi^0(\Lambda  \pi^0) \bar{\Xi}^0(\bar{\Lambda} \gamma)$ decays}

\author{Peng-Cheng Hong}
\affiliation{Institute of Modern Physics, Fudan University, Shanghai 200433, People's Republic of China}
\author{Rong-Gang Ping}
\email[]{pingrg@ihep.ac.cn}
\affiliation{Institute of High Energy Physics, Beijing 100049, People's Republic of China}
\author{He Li}
\affiliation{Department of Modern Physics, University of Science and Technology of China,
	No.96, Jinzhai Road, Hefei, China}
\author{Xiao-Rong Zhou}
\affiliation{Department of Modern Physics, University of Science and Technology of China,
	No.96, Jinzhai Road, Hefei, China}
\author{Tao Luo}
\email[]{luot@fudan.edu.cn}
\affiliation{Institute of Modern Physics, Fudan University, Shanghai 200433, People's Republic of China}

\begin{abstract}
The measurement of decay parameters is one of the important goals of particle physics experiments, and the measurement serves as a probe to search for evidence of CP violation in baryonic decays.
The experimental results will help advance existing theoretical research and establish new experimental objectives.
In this paper, we formulate the asymmetric parameters that characterize parity violation, and then derive formulas for the measurement of CP violation. The formulae for the joint angular distribution of the full decay chain as well as the polarization observable of $\Xi ^ 0 $, $\bar { \Xi } ^ 0 $, $\Lambda $ and $\bar { \Lambda } $ are also provided for experiments. Lastly, we evaluated the sensitivity of two asymmetric parameters: $\alpha _ { \Xi ^ 0 \to \Lambda \pi ^ 0 } $ (abbreviated as $\alpha _ { \Xi ^ 0 } $) and $\alpha _ { \bar { \Xi } ^ 0 \to \bar { \Lambda } \gamma } $ (abbreviated as $\alpha_ {\bar{\Xi}^0 } $) for future experimental measurement.
\end{abstract}

\maketitle


\section{Introduction}\label{intro}
The decay parameters are the key to connect theoretical models with experimental studies. Two-body decays can provide a clean environment to study the properties of baryons, such as polarization and decay parameters, so that theoretical models such as perturbative QCD can be verified. CP violation (CPV) is observed in the $K^0, B^0 $and $D^0 $meson decays \cite{CPVinK, CPVinB1, CPVinB2, CPVinD} and the experimental results are all consistent with the Standard Model predictions.
In the baryonic decay, the magnitude of CPV is predicted only $10 ^ { -4 } \sim 10 ^ { -5 }$ with Standard Model(SM), but can be $10 ^{ -3} $ in some new physics models, such as CP in Refs.~\cite{CP in Baryon 03,CP in Baryon 04,CP in Baryon 05,CP in Baryon 06,CP in Baryon 07,CP in Baryon 08}. However, it is still not large enough to understand the asymmetry of the matter and anti-matter in the universe. Therefore, it is important to expand the sources of CPV, especially in the baryonic sector.

The BESIII at BEPCII has accumulated about 10 billion $J/\psi$ mesons, and a large statistic of hyperon-antihyperon pairs produced from Jpsi decays.
$\ee$ collision experiment has a natural advantage over $pp$ collision or target experiments in measuring high accuracy due to its lower background.
An important work related to our analysis has been done and published in Nature~\cite{nature} with a much more accuracy improvement in measurement.
Searching for evidence of CPV at BESIII experiments remains promising and deserves us to dig deeper in data analysis.

Particle Data Group(PDG) provides an evaluation of $\alpha_{\Xi^0 \to \Lambda \pi^0}=-0.349 \pm 0.009$ via dividing $\alpha({\Xi^0})\alpha_{-}({\Lambda})$ by a current average $\alpha_{-}({\Lambda})$ according to the measurements in recent years \cite{alpha_Xi}. And $\alpha_{\Xi^0\to \Lambda \gamma}$ is equal to $-0.704 \pm0.019_{stat} \pm 0.064_{syst}$ based on the latest result meausred by NA481 Collaboration with a 52000 events data sample \cite{alpha_Xibar}, from which a $2.70\%$ statistical uncertainty can be reached in the radiative decay $\Xi^0\to \Lambda \gamma$.
Once more data sample is available, and simultaneous measurements can be made on its conjugate decay channel, more precise asymmetric parameters can be obtained.
This severs a probe to search for the evidence of CPV in these decays and can help us better understand the mechanism of CPV in baryons.

In this paper, we formulate the observables of parity violation and CPV  as proposed in Ref~\cite{HelicityMethod} in Sec. \ref{PV}.
Different from traditional definition raised by T.~D.~Lee and C.~N.~Yang \cite{LY} with partial wave amplitudes, we use helicity formalism to present these asymmetric parameters.
The method is friendly for experimental physicists to estimate or predict these properties.
We formulate the joint spin density matrix(SDM) of baryon pairs $\Xi^0\bar{\Xi}^0$ and $\Lambda \bar{\Lambda}$ in Sec.~\ref{SDM}.
An sensitivity estimation on the asymmetric parameters of parity violation is performed in Sec.~\ref{estimation}, which provides a reference for accurate measurement on these decay channels in future experiments with high statistic.

\section{asymmetric parameters}\label{PV}
In the two-body decays with parity conservation, the helicity amplitudes satisfy the following symmetry,
\setlength\abovedisplayskip{3pt}
\setlength\belowdisplayskip{3pt}
\begin{eqnarray}\label{Amp for PC}
		A^J_{\lambda_1,\lambda_2}=\eta\eta_1\eta_2(-1)^{J-s_1-s_2} A^J_{-\lambda_1, -\lambda_2},
\end{eqnarray}
where $J, s_1$ and $s_2$ are the spins of the mother particle and the two daughter particles, respectively. $\lambda, \lambda_1,$ and $\lambda_2$ are their helicity values and $\eta, \eta_1$ and $\eta_2$ are their intrinsic parity values, respectively. Assuming that the decays listed in Table~\ref{table1} is parity conserved, the corresponding helicity amplitudes, $A,~B,~F,~G$ and $H$, satisfy
\setlength\abovedisplayskip{3pt}
\setlength\belowdisplayskip{3pt}
\begin{eqnarray}\label{SymmetryRelation}
		A_{-\frac{1}{2},-\frac{1}{2}}&=&A_{\frac{1}{2}, \frac{1}{2}},
		A_{-\frac{1}{2},\frac{1}{2}}=A_{\frac{1}{2}, -\frac{1}{2}},\nonumber\\
		B_{\frac{1}{2}}&=&-B_{-\frac{1}{2}},
		F_{1, \frac{1}{2}}=F_{-1, -\frac{1}{2}},\nonumber\\
		H_{\frac{1}{2}}&=&-H_{-\frac{1}{2}},
		G_{\frac{1}{2}}=-G_{-\frac{1}{2}},
\end{eqnarray}
where we write the amplitude $F$ as $F_{\lambda_5,\lambda_4}$ rather than $F_{\lambda_4,\lambda_5}$ to maintain consistency with its definition in Ref.~\cite{alpha_Xibar}.
However, the parity violation in weak decay renders the above equations invalid. Therefore, we define four asymmetric parameters to describe the parity violation,
\setlength\abovedisplayskip{3pt}
\setlength\belowdisplayskip{3pt}
\begin{eqnarray}\label{DefofPV}
		\alpha_{\Xi^0\to \Lambda \pi^0}&=&\frac{|B_{\frac{1}{2}}|^2-|B_{-\frac{1}{2}}|^2}{|B_{\frac{1}{2}}|^2+|B_{-\frac{1}{2}}|^2},\nonumber\\	\alpha_{\bar{\Xi}^0\to \bar{\Lambda} \gamma}&=&\frac{|F_{-1,-\frac{1}{2}}|^2-|F_{1,\frac{1}{2}}|^2}{|F_{-1,-\frac{1}{2}}|^2+|F_{1,\frac{1}{2}}|^2},\nonumber\\
		\alpha_{\Lambda\to p \pi^-}&=&\frac{|H_{\frac{1}{2}}|^2-|H_{-\frac{1}{2}}|^2}{|H_{\frac{1}{2}}|^2+|H_{-\frac{1}{2}}|^2},\nonumber\\
		\alpha_{\bar{\Lambda}\to \bar{p} \pi^+}&=&\frac{|G_{\frac{1}{2}}|^2-|G_{-\frac{1}{2}}|^2}{|G_{\frac{1}{2}}|^2+|G_{-\frac{1}{2}}|^2}.
\end{eqnarray}
The four parameters defined in this paper are numerically equivalent to the partial-wave amplitudes and are consistent with the parameter values provided in the PDG convention.

Further, if CP is conserved in charge conjugate decays,  then the parameters of the conjugate decays have the same absolute values but having inverse sign as the four corresponding parameters above, i.e. $\alpha_{\bar{\Xi}^0\to \bar{\Lambda} \pi^0}=-\alpha_{\Xi^0\to \Lambda \pi^0}$, $\alpha_{\Xi^0\to \Lambda \gamma}=-\alpha_{\bar{\Xi}^0\to \bar{\Lambda} \gamma}$, $\alpha_{\Lambda\to p \pi^-}=-\alpha_{\bar{\Lambda}\to \bar{p} \pi^+}$.
Thus, we can define three observables characterizing the degree of CPV as
\setlength\abovedisplayskip{3pt}
\setlength\belowdisplayskip{3pt}
\begin{eqnarray}\label{CPO}
	A_{CP}^{1}&=&\frac{\alpha_{\bar{\Xi}^0\to \bar{\Lambda} \pi^0}+\alpha_{\Xi^0\to \Lambda \pi^0}}{\alpha_{\bar{\Xi}^0\to \bar{\Lambda} \pi^0}-\alpha_{\Xi^0\to \Lambda \pi^0}},\nonumber\\
	A_{CP}^{2}&=&\frac{\alpha_{\Xi^0\to \Lambda \gamma}+\alpha_{\bar{\Xi}^0\to \bar{\Lambda} \gamma}}{\alpha_{\Xi^0\to \Lambda \gamma}-\alpha_{\bar{\Xi}^0\to \bar{\Lambda} \gamma}},\nonumber\\
	A_{CP}^{3}&=&\frac{\alpha_{\Lambda\to p \pi^-}+\alpha_{\bar{\Lambda}\to \bar{p} \pi^+}}{\alpha_{\Lambda\to p \pi^-}-\alpha_{\bar{\Lambda}\to \bar{p} \pi^+}}
\end{eqnarray}
The non-zero value of the asymmetric parameters in Eq.~\ref{DefofPV} and \ref{CPO} indicates that there is CPV in the decay.
Experimentally, by measuring these conjugate decays separately, we can obtain the corresponding CP violated information. We shorten $\alpha_{\Xi^0\to \Lambda \pi^0}, \alpha_{\bar{\Xi}^0\to \bar{\Lambda} \gamma}, \alpha_{\Lambda\to p \pi^-}, \alpha_{\bar{\Lambda}\to \bar{p} \pi^+}$ as $\alpha_{\Xi^0}, \alpha_{\bar{\Xi}^0}, \alpha_{\Lambda}, \alpha_{\bar{\Lambda}}$ in the following narrative.

When describing parity violation, the helicity amplitudes are more straightforward compared to the covariant amplitude.
The helicity formalism is widely used in experimental measurements~\cite{BESIII:2022udq, BESIII:2022yzp, Belle:2017egg, Belle:2014nuw, LHCb:2021sqa}.
Helicity amplitudes are also used to form the SDM of particles in a decay, and the SDM contains all the dynamical information of the decay.
The angular distributions and polarization are also derived from SDM easily.
Experimentally, the values of these parameters can be determined by fitting the joint angular distribution to the data \cite{extract pars}. The helicity amplitude can be expanded into the $L$-$S$ coupling of the partial wave amplitude through the Clebsch-Gordan coefficient. In view of the convenience of using the helicity amplitude, we use it to analyze the cascade decay.

\section{helicity system}
\begin{figure}[h]
	\includegraphics[width=8cm,height=2cm]{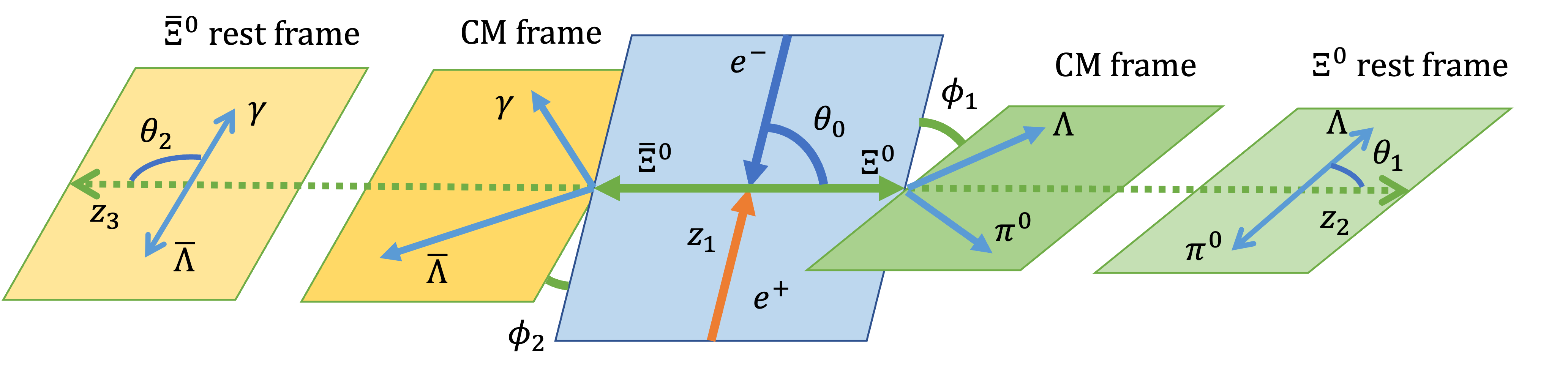}
	\caption{The definition of helicity angles in $J/\psi\to\Xi^0(\Lambda  \pi^0) \bar{\Xi}^0(\bar{\Lambda} \gamma)$ decays}
	\label{Xi_production}
\end{figure}

\begin{figure}[h]
	\includegraphics[width=8cm,height=3cm]{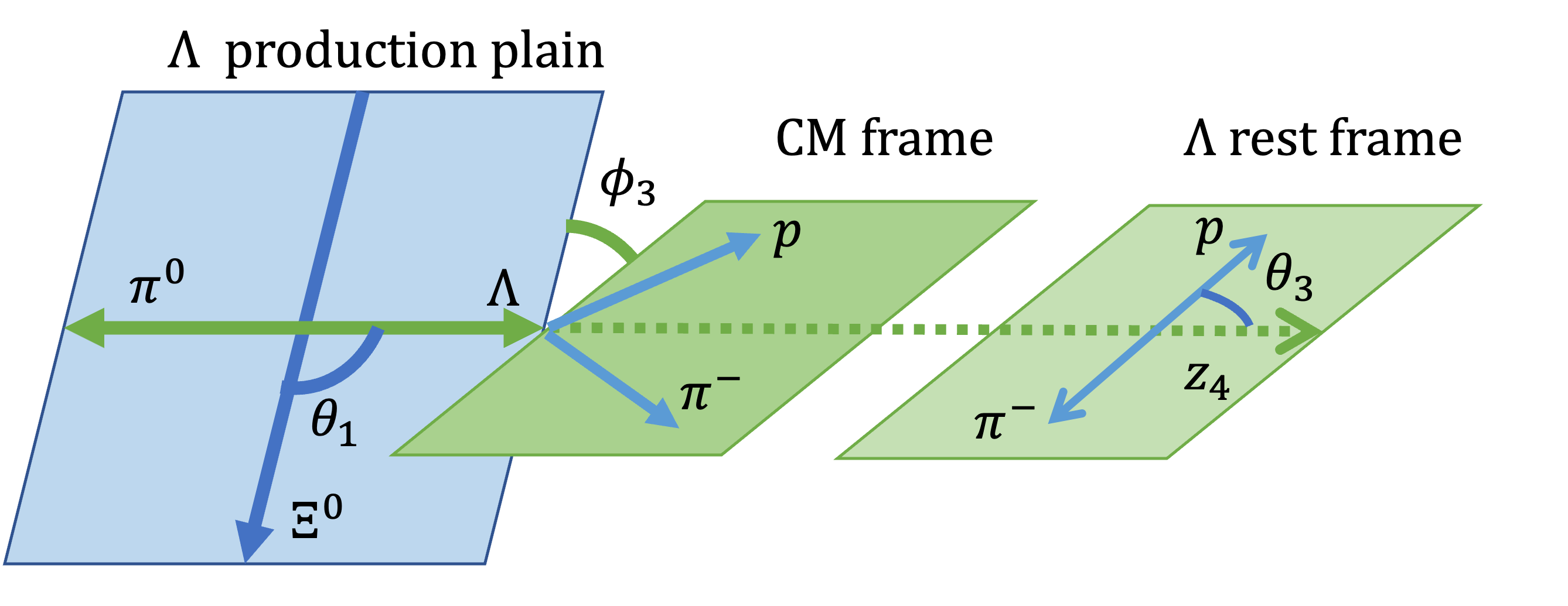}
	\caption{The definition of helicity angles in $\Lambda \to p \pi^0$ decay}
	\label{Lambda_production}
\end{figure}

\begin{figure}[h]
	\includegraphics[width=8cm,height=3cm]{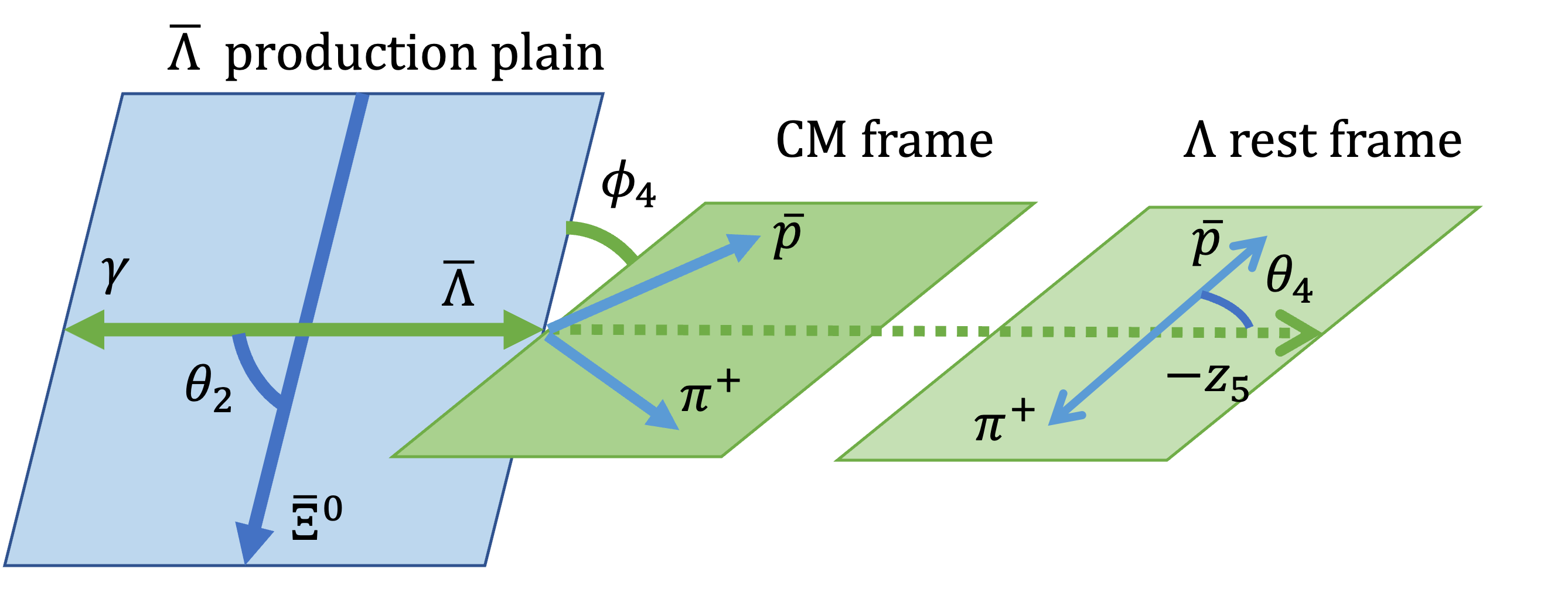}
	\caption{The definition of helicity angles in $\bar{\Lambda} \to \bar{p} \pi^+$ decay}
	\label{Lambdabar_production}
\end{figure}

In this analysis, we use the helicity reference frame to describe the full decay chain. The properties of helicity amplitude can be found in Ref. ~\cite{spinfm}. The helicity angles of the various levels of decay are shown in Fig.~\ref{Xi_production}, Fig.~\ref{Lambda_production} and Fig.~\ref{Lambdabar_production}. The corresponding amplitudes are listed in Table ~\ref{table1}.
\begin{table}[htbp]
	\renewcommand\arraystretch{1.5}
	\caption{Definition of helicity angles and amplitudes in each decay, where $\lambda_i$ indicates the helicity values for the corresponding particles.}
	\label{table1}
	\begin{tabular}{lll}
		\hline\hline
		decay & helicity angle & helicity amplitude \\
		\hline
		$J/\psi\to\Xi^0(\lambda_1)\bar{\Xi}^0(\lambda_2)$
		& ($\theta_0,\phi_0$)
		& $A_{\lambda_1,\lambda_2}$\\
		\hline
		$\Xi^0(\lambda_1^{\prime})\to\Lambda(\lambda_3)\pi^0$ & ($\theta_1,\phi_1$)
		& $B_{\lambda_3}$ \\
		\hline
		$\bar{\Xi}^0(\lambda_2^{\prime})\to\bar{\Lambda}(\lambda_4)\gamma(\lambda_5)$
		& ($\theta_2,\phi_2$)
		& $F_{\lambda_5,\lambda_4}$ \\
		\hline
		$\Lambda(\lambda'_3)\to p(\lambda_6)\pi^-$ & ($\theta_3,\phi_3$)
		& $H_{\lambda_6}$\\
		\hline
		$\bar{\Lambda}(\lambda'_4)\to \bar{p}(\lambda_7) \pi^+$ & ($\theta_4,\phi_4$)
		& $G_{\lambda_7}$\\
		\hline\hline
	\end{tabular}
\end{table}

In this section, we specify that the momentum $p$ with the superscript $L$ represents the momentum in the laboratory system and the momentum without the superscript represents the momentum after the boost operation to the rest frame of its mother particel.
In experiments, the momenta $ \vec{p}^{L}_{{\Lambda}}$ by $\vec{p}^{L}_p$ and $\vec{p}^{L}_{\pi^-}$ are reconstructed from the detection information.
Then we boost $\vec{p}^{L}_p$ and $\vec{p}^{L}_{\pi^-}$ to the $\Lambda$ rest frame, $\theta_3$ describe the angle between $\vec{p}_p$(in the $\Lambda$ rest frame) and $z_4$ axis.
The angle between $\Lambda$ production plane and its decay plane is defined as $\phi_3$.
As for other helicity angles $(\theta_i, \phi_i), (i=0,1,2,4)$, they can be calculated by the same operation as illustration Fig.~\ref{Xi_production} and Fig.~\ref{Lambda_production}.
Notice that the $z_5$ axis along the opposite direction of  $\vec{p}_{\bar{\Lambda}}$ and $\theta_2$ describe the angle between $\vec{p}_{\gamma}$(in the $\bar{\Xi}^0$ rest frame) and $z_3$ axis.
Here, we list all the helicity angle expressions,
\setlength\abovedisplayskip{3pt}
\setlength\belowdisplayskip{3pt}
\begin{eqnarray}\label{angles calculate}
	\theta_0 &=& \arccos({\frac{\vec{p}_{{\Xi^0}} \cdot \vec{p}_{{e^+}}}{|\vec{p}_{{\Xi^0}}| \cdot |\vec{p}_{{e^+}}|}}),~~~~~~~
	\phi_0 = 0,\nonumber\\
	\theta_1 &=& \arccos({\frac{\vec{p}_{{\Xi^0}} \cdot \vec{p}_{{\Lambda}}}{|\vec{p}_{{\Xi^0}}| \cdot |\vec{p}_{{\Lambda}}|}}),
	\phi_1 =\arccos({ |\vec{n}_{J/\psi} \cdot \vec{n}_{\Xi^0}|}),\nonumber\\
	\theta_2 &=& \arccos({\frac{\vec{p}_{\bar{\Xi}^0} \cdot \vec{p}_{\gamma}}{|\vec{p}_{\bar{\Xi}^0}| \cdot |\vec{p}_{\gamma}|}}),
	\phi_2 = \arccos({ |\vec{n}_{J/\psi} \cdot \vec{n}_{\bar{\Xi}^0}|}),\nonumber\\
	\theta_3 &=&\arccos({ \frac{\vec{p}_{{\Lambda}} \cdot \vec{p}_{{p}}}{|\vec{p}_{{\Lambda}}| \cdot |\vec{p}_{{p}}|}}),
	\phi_3 = \arccos({|\vec{n}_{\Xi^0} \cdot \vec{n}_{\Lambda}|}),\nonumber\\
	\theta_4 &=& \arccos({\frac{\vec{p}_{\bar{\Lambda}} \cdot \vec{p}_{\bar{p}}}{|\vec{p}_{\bar{\Lambda}}| \cdot |\vec{p}_{\bar{p}}|}}),
	\phi_4 =  \arccos({|\vec{n}_{\bar{\Xi}^0} \cdot \vec{n}_{\bar{\Lambda}}|}),
\end{eqnarray}
where the unit vectors $\vec{n}_{m}$ in the rest frame of $m$ decay plane are defined with the momenta of those particles as
\setlength\abovedisplayskip{3pt}
\setlength\belowdisplayskip{3pt}
\begin{eqnarray}
	\vec{n}_{J/\psi} &=& \frac{\vec{p}_{e^+}\times{\vec{p}_{\Xi^0}}}{|\vec{p}_{e^+}| \cdot |\vec{p}_{\Xi^0}| \cdot \sin\theta_0},
	\vec{n}_{\Xi^0} = \frac{\vec{p}_{\Xi^0}\times{\vec{p}_{\Lambda}}}{|\vec{p}_{\Xi^0}| \cdot |\vec{p}_{\Lambda}| \cdot \sin\theta_1},\nonumber\\
	\vec{n}_{\bar{\Xi^0}} &=& \frac{\vec{p}_{\bar{\Xi}^0}\times{\vec{p}_{\gamma}}}{|\vec{p}_{\bar{\Xi}^0}| \cdot |\vec{p}_{\gamma}| \cdot \sin\theta_2},
	\vec{n}_{\Lambda} = \frac{\vec{p}_{\Lambda}\times{\vec{p}_{p}}}{|\vec{p}_{\Lambda}| \cdot |\vec{p}_{p}| \cdot \sin\theta_3},\nonumber\\
	\vec{n}_{\bar{\Lambda}} &=& \frac{\vec{p}_{\bar{\Lambda}}\times{\vec{p}_{\bar{p}}}}{|\vec{p}_{\bar{\Lambda}}| \cdot |\vec{p}_{\bar{p}}| \cdot \sin\theta_4}.
\end{eqnarray}

\section{spin density matrix and angular distribution}\label{SDM}
Since the SDM contains all the dynamical information in the decay, we first calculate the SDM of baryons in each step of decay, and then derive the angular distributions and the expression to present baryon polarization~\cite{spinfm, Spin}.

\subsection{$J/\psi \to \Xi^0 \bar{\Xi}^0$}
For a spin-$\frac{1}{2}$ particle like $\Xi^0$, the SDM can be expressed as
\setlength\abovedisplayskip{3pt}
\setlength\belowdisplayskip{3pt}
\begin{eqnarray}
	\rho^{\Xi^0}=
	\left(
	\begin{array}{cc}
		\rho^{\Xi^0}_{\frac{1}{2}, \frac{1}{2}} & \rho^{\Xi^0}_{\frac{1}{2}, -\frac{1}{2}} \\
		\rho^{\Xi^0}_{-\frac{1}{2}, \frac{1}{2}} & \rho^{\Xi^0}_{-\frac{1}{2}, -\frac{1}{2}}
	\end{array}
	\right).
\end{eqnarray}
The joint SDM of $\Xi^0 \bar{\Xi}^0$ can be constructed in the form of $\rho^{\Xi^0} \otimes \rho^{\bar{\Xi}^0}$, and its elements can be directly calculated as
\setlength\abovedisplayskip{3pt}
\setlength\belowdisplayskip{3pt}
\begin{eqnarray}\label{ele of rhoXX}
	\rho^{\Xi^0\bar{\Xi}^0}_{\lambda_1,\lambda_2,\lambda_1',\lambda_2'}&\propto&\sum_{\lambda,\lambda'}\rho_{\lambda,\lambda'}^{\psi}D^{J*}_{\lambda,\lambda_1-\lambda_2}(\phi_0,\theta_0,0)\nonumber\\
	&\times& D^{J}_{\lambda',\lambda_1'-\lambda_2'}(\phi_0,\theta_0,0)A_{\lambda_1,\lambda_2}A^*_{\lambda'_1,\lambda'_2},
\end{eqnarray}
where the SDM of $J/\psi$ produced from $e^+e^-$ annihilation can be described as $\rho_{\lambda,\lambda'}^{\psi}=\frac{1}{2}\mathrm{diag}\{1,0,1\}$ \cite{PolaInChicj}, and $D^{J}_{\lambda_i,\lambda_k}(\phi_0,\theta_0,0)$ is the Wigner-D function.
Since the $J/\psi$ decaying into $\Xi^0\bar{\Xi}^0$ via strong interactions conserves the parity, the helicity amplitudes satisfy the equations listed in Eq.~(\ref{SymmetryRelation}) i.e. $A_{-\frac{1}{2},-\frac{1}{2}}=A_{\frac{1}{2}, \frac{1}{2}}, A_{-\frac{1}{2},\frac{1}{2}}=A_{\frac{1}{2}, -\frac{1}{2}}$.
The angular distribution of $J/\psi \to \Xi^0 \bar{\Xi}^0$ can be expressed as
\begin{eqnarray}
		I(\theta_0) &\propto& \mathrm{Tr}[\rho^{\Xi^0\bar{\Xi}^0}]=|A_{\frac{1}{2},\frac{1}{2}}|^2 \sin ^2\theta _0\nonumber\\
		&+&\frac{1}{4} |A_{\frac{1}{2},-\frac{1}{2}}|^2 \left(\cos 2 \theta _0+3\right).
\end{eqnarray}
If we choose
\begin{eqnarray}\label{alpha psi}
		\alpha_\psi=\frac{{|A_{\frac{1}{2},-\frac{1}{2}}|^2-2|A_{\frac{1}{2},\frac{1}{2}}|^2}}{{|A_{\frac{1}{2},-\frac{1}{2}}|^2+2|A_{\frac{1}{2},\frac{1}{2}}|^2}},
\end{eqnarray}
then the angular distribution can be reduced to the formula commonly used in experiments
\begin{eqnarray}
		I(\theta_0) \propto 1+\alpha _\psi\cos ^2\theta _0,
\end{eqnarray}
where $\alpha_{\psi }$ is the angular distribution parameter.

On the other hand, the joint SDM of $\Xi^0 \bar{\Xi}^0$ can also be expressed by the real multipole parameters $Q^{1}_{i,j}$ as
\setlength\abovedisplayskip{0pt}
\setlength\belowdisplayskip{3pt}
\begin{eqnarray}\label{Q1ij expand}
	\rho^{\Xi^0\bar{\Xi}^0}=\frac{Q^{1}_{0,0}}{4}[I+\sum_{\overline{i,j=0}}^{3}Q^{1}_{i,j}\sigma^{\Xi^0}_i\otimes\sigma^{\bar{\Xi}^0}_j],
\end{eqnarray}
where the superscript of $Q^{1}_{i,j}$ is used to distinguish from the parameters $Q^{2}_{i,j}$ used in Eq.~(\ref{Q2ij expand}), ${I}$ is a $4 \times 4$ identity matrix and $\sigma$ is Pauli matrix \cite{PolaInChicj}.
Here, $\sigma_i$ or $\sigma_j~(i,j = 1, 2, 3)$ correspond to $\sigma_x, \sigma_y, \sigma_z$, and denote by $i$ or $j = 0 $ the $2\times 2 $ identity matrix, $\overline{i,j=0}$ means they cannot be 0 at the same time.
$Q^1_{i, j}$ can be calculated by $Q^{1}_{0,0}=\mathrm{Tr} \rho^{\Xi^0\bar{\Xi}^0},  Q^{1}_{0,0}Q^{1}_{i,j}=\mathrm{Tr}[\sigma_i\otimes\sigma_j\cdot\	\rho^{\Xi^0\bar{\Xi}^0}]$.
In this way, the multipole parameters $Q^1_{i, j}$ can be expressed with the helicity amplitudes as listed in Eq.~(\ref{Q1expression}).

For the decay $J/\psi \to \Xi^0 \bar{\Xi}^0$, $Q^1_{0, 0}$ stands for the unpolarized decay rate.
The degree of $\Xi^0$ linear polarization can be expressed as $\mathcal{P}_x^{\Xi^0}=Q^{1}_{1,0},\mathcal{P}_y^{\Xi^0}=Q^{1}_{2,0}$ and the longitudinal polarization $\mathcal{P}_z^{\Xi^0}=Q^{1}_{3,0}$.
For the $\bar{\Xi}^0$, they are $\mathcal{P}_x^{\bar{\Xi}^0}=Q^{1}_{0,1},\mathcal{P}_y^{\bar{\Xi}^0}=Q^{1}_{0,2}$ and $\mathcal{P}_z^{\bar{\Xi}^0}=Q^{1}_{0,3}$. Since the parity conserves in the $J/\psi$ decay, we have the polarization expressions as
\setlength\abovedisplayskip{3pt}
\setlength\belowdisplayskip{3pt}
\begin{eqnarray}\label{XiXIbarPola}
	\mathcal{P}^{\Xi^0}_x &=& -\mathcal{P}^{\bar{\Xi}^0}_x=0,~~~~~~
	\mathcal{P}^{\Xi^0}_z = -\mathcal{P}^{\bar{\Xi}^0}_z=0,\nonumber\\
	\mathcal{P}^{\Xi^0}_y &=& -\mathcal{P}^{\bar{\Xi}^0}_y
	= \frac{\sqrt{1-\alpha_{\psi}^2} \sin{2\theta_0} \sin{\Delta_a}}{2(1+\alpha_{\psi} \cos^2{\theta_0})},
\end{eqnarray}
where $\Delta _a = \xi_{\frac{1}{2},-\frac{1}{2}}-\xi_{\frac{1}{2},\frac{1}{2}}$ is the phase difference of the two amplitudes $A_{{1\over2},-{1\over2}},A_{{1\over2},{1\over2}}$. Obviously, whether the transverse polarization exists or not depends on the phase angle difference $\Delta _a$.

\subsection{$\Xi^0(\bar{\Xi}^0) \to \Lambda \pi^0 (\bar{\Lambda} \gamma)$}
In these two decays $\Xi^0 \to \Lambda \pi^0$ and $\bar{\Xi}^0 \to \bar{\Lambda} \gamma$, the parity violation can be revealed by the study of the angular distribution of the decaying particle or by the measurement of the polarization. The joint angular distribution $I(\theta_0,\theta_1,\phi_1,\theta_2,\phi_2)$ of this decay can be calculated by the joint SDM of $\Xi^0 \bar{\Xi}^0$ as
\setlength\abovedisplayskip{0pt}
\setlength\belowdisplayskip{3pt}
\begin{eqnarray}
	I(\theta_0, \theta_1, \phi_1, \theta_2, \phi_2)
	&\propto&\sum_{\lambda_i,\lambda'_i}\rho^{\Xi^0\bar{\Xi}^0}_{\lambda_1,\lambda_2;\lambda'_1,\lambda'_2}
	D^{\frac{1}{2}*}_{\lambda_1,\lambda_3}(\theta_1,\phi_1)\nonumber\\
	&\times&D^{\frac{1}{2}}_{\lambda'_1,\lambda_3}(\theta_1,\phi_1)
	D^{\frac{1}{2}*}_{\lambda_2,\lambda_5-\lambda_4}(\theta_2,\phi_2)\nonumber\\
	&\times&D^{\frac{1}{2}}_{\lambda'_2,\lambda_5-\lambda_4}(\theta_2,\phi_2)
	B_{\lambda_3}B^{*}_{\lambda_3}\nonumber\\
	&\times&
	F_{\lambda_5,\lambda_4}F^{*}_{\lambda_5,\lambda_4}
\end{eqnarray}
where the summation is taken over all involved helicities $\lambda_i$ and $\lambda'_i,(i=1,2,3,4,5)$.
We factor out the constant term and then simplify the angular distribution as
\setlength\abovedisplayskip{0pt}
\setlength\belowdisplayskip{3pt}
\begin{eqnarray}\label{JointAng02}
	I(\theta_0, \theta_1, \phi_1, \theta_2, \phi_2)
	&\propto&
	1+\alpha _{\psi } \cos ^2\theta _0\nonumber\\
	&+&\sqrt{1-\alpha _{\psi }^2} \sin \theta _0 \cos \theta _0 \nonumber\\
	&\times&\left\{\right. \sin \theta _2 \sin \phi _2 \alpha _{\bar{\Xi }^0} \sin \Delta _a\nonumber\\
	&+&\alpha _{\Xi^0} [\alpha _{\bar{\Xi }^0} \cos \Delta _a (\sin \theta _2 \cos \theta _1 \cos \phi _2\nonumber\\
	&-&\sin \theta _1 \cos \theta _2 \cos \phi _1)\nonumber\\
	&+&\sin \theta _1 \sin \phi _1 \sin \Delta _a]\left.\right\}\nonumber\\
	&-&\alpha _{\Xi^0} \alpha _{\bar{\Xi}^0} [-\cos \theta _1 \cos \theta _2 (\alpha _{\psi }+\cos ^2\theta _0)\nonumber\\
	&+&\alpha _{\psi } \sin \theta _1 \sin \theta _2 \sin ^2\theta _0 \sin \phi _1 \sin \phi _2\nonumber\\
	&+&\sin \theta _1 \sin \theta _2 \sin ^2\theta _0 \cos \phi _1 \cos \phi _2],
\end{eqnarray}
where $\alpha_{\Xi^0}$ and $\alpha_{\bar{\Xi}^0}$ measure parity violation.

In order to simplify the calculations of next decay chain, we adopt the $\Xi$ decay matrices to describe the joint angular distribution $I(\theta_0, \theta_1, \phi_1, \theta_2, \phi_2)$ \cite{PolaInChicj}, i.e.,
\setlength\abovedisplayskip{2pt}
\setlength\belowdisplayskip{3pt}
\begin{eqnarray}\label{trace1}
	I(\theta_0, \theta_1, \phi_1, \theta_2, \phi_2)\propto
	\mathrm{Tr}[\rho^{\Xi^0\bar{\Xi}^0}\cdot(M^{\Xi^0}\otimes M^{\bar{\Xi}^0})^T],
\end{eqnarray}
where $M^{\Xi^0}(M^{\bar{\Xi}^0})$ is the decay matrix of $\Xi^0(\bar{\Xi}^0)$, and its elements can be expressed as
\setlength\abovedisplayskip{0pt}
\setlength\belowdisplayskip{3pt}
\begin{eqnarray}\label{ele of decay matrix}
	|M_{\lambda,\lambda'}|^{2}&=&\sum_{\lambda_1,\lambda_2} D^{J*}_{\lambda,\lambda_1-\lambda_2} (\alpha,\beta,\gamma)
	D^{J}_{\lambda',\lambda_1-\lambda_2} (\alpha,\beta,\gamma)\nonumber\\
	&\times&A_{\lambda_1,\lambda_2} A^{*}_{\lambda_1,\lambda_2},
\end{eqnarray}
where $J$ and $\lambda$ represent the spin and helicity of the mother particle, and $\lambda_1, \lambda_2$ are the helicities of the daughter particles, respectively. $A_{\lambda_1,\lambda_2}$ is the helicity amplitude and $(\alpha,\beta,\gamma)$ corresponds to the helicity angles in the decay.
For $\Xi^0$ and $\bar{\Xi}^0$, they are $(\phi_1, \theta_1, 0)$ and $(\phi_2, \theta_2, 0)$, then we have
\setlength\abovedisplayskip{2pt}
\setlength\belowdisplayskip{3pt}
\begin{eqnarray}
	M^{\Xi^0}&=&\frac{1}{2}
	\left(
	\begin{array}{cc}
		1+\alpha_{\Xi^0}\cos\theta_1 &  e^{i\phi_1}\alpha_{\Xi^0}\sin\theta_1 \\
		e^{-i\phi_1}\alpha_{\Xi^0}\sin\theta_1  &  1-\alpha_{\Xi^0}\cos\theta_1
	\end{array}
	\right),\nonumber\\
	M^{\bar{\Xi}^0}&=&\frac{1}{2}
	\left(
	\begin{array}{cc}
		1-\alpha_{\bar{\Xi}^0}\cos\theta_2 &  -e^{i\phi_2}\alpha_{\bar{\Xi}^0}\sin\theta_2 \\
		-e^{-i\phi_2}\alpha_{\bar{\Xi}^0}\sin\theta_2  &  1+\alpha_{\bar{\Xi}^0}\cos\theta_2
	\end{array}
	\right).
\end{eqnarray}
The joint angular distribution $I(\theta_0, \theta_1, \phi_1, \theta_2, \phi_2)$ in this form is expressed as
\setlength\abovedisplayskip{2pt}
\setlength\belowdisplayskip{3pt}
\begin{eqnarray}\label{JointAng03}
	I(\theta_0, \theta_1, \phi_1, \theta_2, \phi_2)
	&\propto&	Q^{1}_{0,0} \left\{\right.1+Q^{1}_{2,0} \alpha _{\Xi^0 } \sin \theta _1 \sin \phi _1+\alpha _{\bar{\Xi }^0}\nonumber\\
	&\times&[-\alpha _{\Xi^0 } (Q^{1}_{1,1} \sin \theta _1 \sin \theta _2 \cos \phi _1 \cos \phi _2\nonumber\\
	&+&Q^{1}_{1,3} \sin \theta _1 \cos \theta _2 \cos \phi _1\nonumber\\
	&+&\sin \theta _2 (Q^{1}_{2,2} \sin \theta _1 \sin \phi _1 \sin \phi _2 \nonumber\\
	&+&Q^{1}_{3,1} \cos \theta _1 \cos \phi _2)+Q^{1}_{3,3} \cos \theta _1 \cos \theta _2)\nonumber\\
	&-&Q^{1}_{0,2} \sin \theta _2 \sin \phi _2] \left.\right\}.
\end{eqnarray}
If we substitute the $Q^1_{i, j}$ with Eq.~(\ref{Q1expression}), we can see that it is consistent with Eq.~(\ref{JointAng02}).
Furthermore, $I(\theta_0, \theta_1, \phi_1, \theta_2, \phi_2)$ can be written as
\setlength\abovedisplayskip{0pt}
\setlength\belowdisplayskip{3pt}
\begin{eqnarray}\label{JointAng04}
	I(\theta_0\sim \phi_2)
	\propto
	Q^{1}_{0,0}+T^1_1 \alpha_{\Xi^0}+
	\bar{T}^1_1 \alpha_{\bar{\Xi}^0}
	+
	T^1_2 \alpha_{\Xi^0}\alpha_{\bar{\Xi}^0},
\end{eqnarray}
where the superscript of $T^1_{i}$ is used to distinguish from the parameters $T^2_{i}$ used in Eq.~(\ref{T matrix 2}).
$T^1_1$ and $\bar{T}^1_1$ measure the transverse polarization information of $\Xi^0$ and $\bar{\Xi}^0$, respectively.
$T^1_2$ measures the $\Xi^0 \bar{\Xi}^0$ spin correlations.
They are
\setlength\abovedisplayskip{3pt}
\setlength\belowdisplayskip{3pt}
\begin{eqnarray}
	T^1_1 &=& \sin \theta _1 \sin \phi _1Q^{1}_{0,0}Q^{1}_{2,0},\nonumber\\
	\bar{T}^1_1 &=& -\sin\theta_2 \sin\phi_2 Q^{1}_{0,0}Q^{1}_{0,2},\nonumber\\
	T^1_2 &=& -Q^{1}_{0,0}(Q^{1}_{2,2} \sin \theta _1 \sin \theta _2 \sin \phi _1 \sin \phi _2\nonumber\\
	&+&Q^{1}_{1,1} \sin  \theta _1 \sin  \theta _2 \cos  \phi _1 \cos \phi _2\nonumber\\
	&+& Q^{1}_{1,3} \sin  \theta _1 \cos  \theta _2 \cos \phi _1
	+Q^{1}_{3,1} \sin  \theta _2 \cos  \theta _1 \cos  \phi _2\nonumber\\
	&+&Q^{1}_{3,3} \cos  \theta _1 \cos  \theta _2).
\end{eqnarray}

\subsection{$\Lambda (\bar{\Lambda}) \to p \pi^- (\bar{p} \pi^+)$}
The parameters to measure the parity violation in the weak decays $\Lambda \to p \pi^-$ and $\bar{\Lambda} \to \bar{p} \pi^+$ have been defined in Eq.~(\ref{DefofPV}).
The elements of the joint SDM of $\Lambda \bar{\Lambda}$ can be given by the joint SDM of $\Xi^0 \bar{\Xi}^0$ as
\setlength\abovedisplayskip{3pt}
\setlength\belowdisplayskip{3pt}
\begin{eqnarray}
	\rho^{\llb}_{\lambda_3,\lambda_4,\lambda'_3,\lambda'_4}
	&\propto&
	\sum_{\lambda_1,\lambda_2,\lambda'_1,\lambda'_2,\lambda_5}
	\rho^{\Xi^0\bar{\Xi}^0}_{\lambda_1,\lambda_2;\lambda'_1,\lambda'_2}
	D^{\frac{1}{2}*}_{\lambda_1,\lambda_3}(\theta_1,\phi_1)\nonumber\\
	&\times&
	D^{\frac{1}{2}}_{\lambda'_1,\lambda'_3}(\theta_1,\phi_1)B_{\lambda_3}B^{*}_{\lambda'_3}
	D^{\frac{1}{2}*}_{\lambda_2,\lambda_5-\lambda_4}(\theta_2,\phi_2)\nonumber\\
	&\times&
	D^{\frac{1}{2}}_{\lambda'_2,\lambda_5-\lambda'_4}(\theta_2,\phi_2)
	F_{\lambda_5,\lambda_4}F^{*}_{\lambda_5,\lambda'_4},
\end{eqnarray}
and the specific expressions are shown in Eq.~(\ref{rhoLL}).
Here we use the SDM of $\Xi^0 \bar{\Xi}^0$ with the $Q^{1}_{i,j}$ parameters.
Also it can be calculated by a direct product of $\rho^{\Lambda}$ and $\rho^{\bar{\Lambda}}$ as well.

Analogous to Eq.~(\ref{Q1ij expand}), we get the $\Lambda\bar\Lambda$ joint SDM $\rho^{\Lambda\bar{\Lambda}}$ with multipole parameters $Q^2_{i, j}$ as
\setlength\abovedisplayskip{0pt}
\setlength\belowdisplayskip{3pt}
\begin{eqnarray}\label{Q2ij expand}
		\rho^{\Lambda\bar{\Lambda}}=\frac{Q^{2}_{0,0}}{4}[I+\sum_{\overline{i,j=0}}^{3}Q^{2}_{i,j}\sigma^{\Lambda}_i\otimes\sigma^{\bar{\Lambda}}_j],
\end{eqnarray}
where $Q^2_{i, j}$ stands for the polarizations and spin correlations of $\Lambda \bar{\Lambda}$.
As same as the situation of $\Xi^0 \bar{\Xi}^0$, the polarization of $\Lambda$ and $\bar{\Lambda}$ can be expressed as
\begin{eqnarray}
			\mathcal{P}^{\Lambda}_x &=&Q^2_{1, 0}, ~~~\mathcal{P}^{\Lambda}_y =Q^2_{2, 0},  ~~~	\mathcal{P}^{\Lambda}_z =Q^2_{3, 0} ,\nonumber\\
			\mathcal{P}^{\bar{\Lambda}}_x&=&Q^2_{0, 1},~~~\mathcal{P}^{\bar{\Lambda}}_y=Q^2_{0, 2}, ~~~\mathcal{P}^{\bar{\Lambda}}_z=Q^2_{0, 3},
\end{eqnarray}
The expressions of $Q^2_{i, j}$ are listed in Eq.~(\ref{Q2expression}).

Using Eq.~(\ref{ele of decay matrix}), we get the decay matrices of $\Lambda$ and $\bar{\Lambda}$ as
\setlength\abovedisplayskip{3pt}
\setlength\belowdisplayskip{3pt}
\begin{eqnarray}
	M^{\Lambda}&=&\frac{1}{2}
	\left(
	\begin{array}{cc}
		1+\alpha_{\Lambda}\cos\theta_3 &  e^{i\phi_3}\alpha_{\Lambda}\sin\theta_3 \\
		e^{-i\phi_3}\alpha_{\Lambda}\sin\theta_3  &
		1-\alpha_{\Lambda}\cos\theta_3
	\end{array}
	\right),\nonumber\\
	M^{\bar{\Lambda}}
	&=&\frac{1}{2}
	\left(
	\begin{array}{cc}
		1+\alpha_{\bar{\Lambda}}\cos\theta_4 &  e^{i\phi_4}\alpha_{\bar{\Lambda}}\sin\theta_4 \\
		e^{-i\phi_4}\alpha_{\bar{\Lambda}}\sin\theta_4  &
		1-\alpha_{\bar{\Lambda}}\cos\theta_4
	\end{array}
	\right).
\end{eqnarray}
Combined with Eq.~(\ref{trace1}), the joint angular distribution $I(\theta_0,\theta_1,\phi_1,\theta_2,\phi_2,\theta_3,\phi_3,\theta_4,\phi_4)$ at this level can be expressed as
\setlength\abovedisplayskip{0pt}
\setlength\belowdisplayskip{3pt}
\begin{eqnarray}
	&~&I(\theta_0,\theta_1,\phi_1,\theta_2,\phi_2,\theta_3,\phi_3,\theta_4,\phi_4)\nonumber\\
	&\propto&Q^{2}_{0,0} \left\{\right.1-\cos \theta _4 \alpha _{\bar{\Lambda }} [-\alpha _{\Lambda } (\sin \theta _3 (Q^{2}_{2,3} \sin \phi _3\nonumber\\
	&+&Q^{2}_{1,3} \cos \phi _3) +Q^{2}_{3,3} \cos \theta _3)-Q^{2}_{0,3}]\nonumber\\
	&+&\alpha _{\Lambda } [\sin \theta _3 (Q^{2}_{2,0} \sin \phi _3 +Q^{2}_{1,0} \cos \phi _3)\nonumber\\
	&+&Q^{2}_{3,0} \cos \theta _3]\left.\right\}.
\end{eqnarray}
Analogue to Eq.~(\ref{JointAng04}), it can be simplified as
\setlength\abovedisplayskip{3pt}
\setlength\belowdisplayskip{3pt}
\begin{eqnarray}\label{T matrix 2}
	&~&I(\theta_0,\theta_1,\phi_1,\theta_2,\phi_2,\theta_3,\phi_3,\theta_4,\phi_4)\nonumber\\
	&\propto&
	Q^{2}_{0,0}
	+T^2_1 \alpha_{\Lambda}
	+
	\bar{T}^2_1 \alpha_{\bar{\Lambda}}+
	T^2_2 \alpha_{\Lambda}\alpha_{\bar{\Lambda}},
\end{eqnarray}
with
\setlength\abovedisplayskip{3pt}
\setlength\belowdisplayskip{3pt}
\begin{eqnarray}
	T^2_1 &=&Q^{2}_{0,0} \sin \theta _3 (Q^{2}_{2,0} \sin \phi _3+Q^{2}_{1,0} \cos \phi _3)\nonumber\\
	&-&Q^{2}_{3,0} \cos \theta _3,\nonumber\\
	\bar{T}^2_1 &=&Q^2_{0,0}Q^2_{0,3} \cos \theta_4 ,\nonumber\\
	T^2_2 &=&Q^{2}_{0,0} \cos \theta_4 [\sin \theta _3 (Q^{2}_{2,3} \sin \phi _3+Q^{2}_{1,3} \cos \phi _3)\nonumber\\
	&+&Q^{2}_{3,3} \cos \theta _3].
\end{eqnarray}
where $T^2_1$ and $\bar{T}^2_1$ respect the transverse polarization information for $\Lambda$ and $\bar{\Lambda}$, respectively, while $T^2_2$ respects the $\Lambda \bar{\Lambda}$ spin correlations, which are similar to the interpretation of Eq.~(\ref{JointAng04}).

\section{sensitivity of asymmetric parameters measurements}\label{estimation}
Sensitivity estimation is the basis of physical experiment design, which reveals the relationship between the measurement accuracy of physical quantities and data statistics. We use the entire decay chain to improve the accuracy of the statistical sensitivity estimate. The results of our calculations show the expected measurement accuracy of these asymmetric parameters in the experiment versus the statistics of the data. The method we use is also applicable to other similar decay processes. For the large-scale experimental devices to be built in the future, such as STCF and CEPC~\cite{STCF1, STCF2, CEPC}, the estimation of sensitivity is urgently needed to guide the data acquisition plan.

In the estimation of sensitivities, we give the normalized angular distribution as
\setlength\abovedisplayskip{3pt}
\setlength\belowdisplayskip{3pt}
\begin{eqnarray}
	\widetilde{\mathcal{W}}=\frac{\mathcal{W}(\theta_0,\theta_1,\phi_1,\theta_2,\phi_2,\theta_3,\phi_3,\theta_4,\phi_4)}{\int\cdot\cdot\cdot\int\mathcal{W}(\cdot\cdot\cdot) \prod_{i=0}^{4} \mathrm{d}\mathrm{cos}\theta_i \prod_{j=1}^{4}\mathrm{d}\phi_j },
\end{eqnarray}
where the different asymmetric parameters used are taking as $\alpha_{\psi}=0.66 \pm 0.03 \pm 0.05,\alpha_{\Lambda}=0.732 \pm 0.014,\alpha_{\bar{\Lambda}}=-0.758 \pm 0.010 \pm 0.007$ according to Refs.~\cite{alpha_psi, alpha_lambda, PDG}. $\alpha_{\bar{\Xi}^0}=0.70 \pm 0.07$ in the hypothesis of CP conservation, and $\alpha_{\Xi^0}=-0.349 \pm 0.009$ as mentioned in Sec.~\ref{intro}.
The phase angle differences arbitrarily take as $\Delta_{a}=\frac{\pi}{3},\Delta_{b}=\frac{\pi}{4},\Delta_{f}=\frac{\pi}{6}$.
We also use other sets of phase angles differences for calculation, and the results show that the sensitivity estimation of the asymmetric parameters for large statistical quantities is not significantly affected.
Here, the maximum likelihood function is defined as
\setlength\abovedisplayskip{3pt}
\setlength\belowdisplayskip{3pt}
\begin{eqnarray}
	L=\prod_{i=1}^{N}\widetilde{\mathcal{W}}(\theta_0,\theta_1,\phi_1,\theta_2,\phi_2,\theta_3,\phi_3,\theta_4,\phi_4),
\end{eqnarray}
where N represents the number of observed events \cite{sensitivity calculation}.
And the variance of the asymmetric parameters, for example, $\alpha_{\Xi^0}$, can be expressed as
\setlength\abovedisplayskip{0pt}
\setlength\belowdisplayskip{3pt}
\begin{eqnarray}
	V^{-1}(\alpha_{\Xi^0})=N\int\frac{1}{\widetilde{\mathcal{W}}}[\frac{\partial	\widetilde{\mathcal{W}}}{\partial\alpha_{\Xi^0}}]^2 \prod_{i=0}^{4} \mathrm{d}\mathrm{cos}\theta_i \prod_{j=1}^{4}\mathrm{d}\phi_j.
\end{eqnarray}
Thus we give the statistical sensitivity of $\alpha_{\Xi^0}$ and $\alpha_{\bar{\Xi}^0}$ as
\setlength\abovedisplayskip{3pt}
\setlength\belowdisplayskip{3pt}
\begin{eqnarray}
	\delta_1=\frac{\sqrt{V(\alpha_{\Xi^0})}}{|\alpha_{\Xi^0}|},~~~\delta_2=\frac{\sqrt{V(\alpha_{\bar{\Xi}^0})}}{|\alpha_{\bar{\Xi}^0}|}.
\end{eqnarray}

\begin{figure}[h]
	\includegraphics[width=8cm,height=5cm]{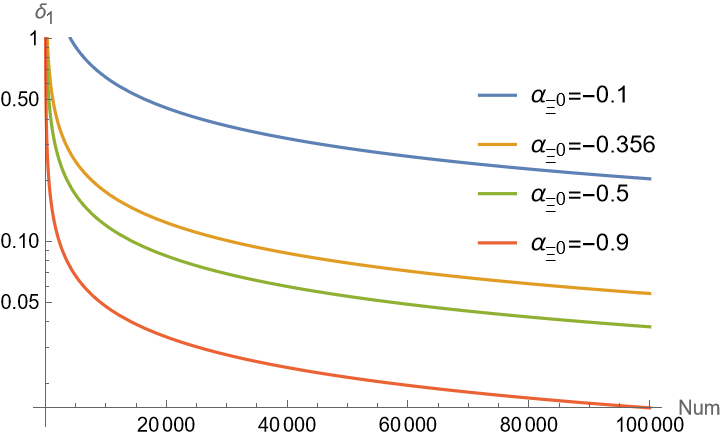}
	\caption{The sensitivity of $\alpha_{\Xi^0}$ relative to observed events N}
	\label{fig aX}
\end{figure}
\begin{figure}[h]
	\includegraphics[width=8cm,height=5cm]{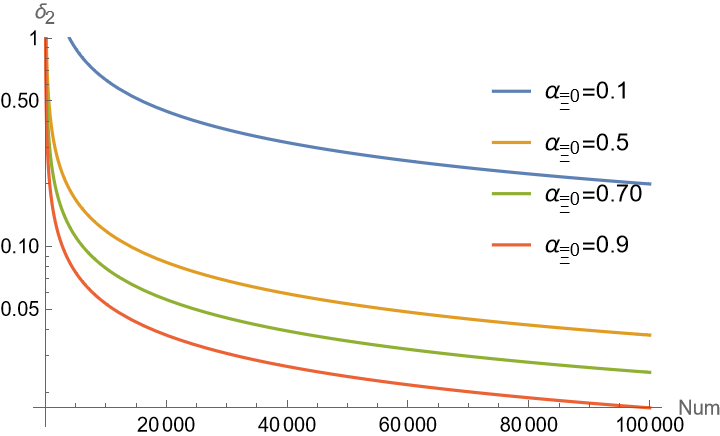}
	\caption{The sensitivity of $\alpha_{\bar{\Xi}^0}$ relative to observed events N}
	\label{fig aXb}
\end{figure}


We take a set of possible $\alpha_{\Xi^0}$ and $\alpha_{\bar{\Xi}^0}$ values to plot the sensitivity as shown in Fig.~\ref{fig aX} and Fig.~\ref{fig aXb}, respectively.
From the figure, we can draw the following conclusions.
First, the larger the absolute value of the asymmetric parameter, the less data is needed to reach the same statistical sensitivity.
This shows that in the case of the same data statistics, a larger asymmetric parameter value means a high measurement accuracy.
Second, from Fig.~\ref{fig aXb} we can see that our predictions on the asymmetric parameter $\alpha_{\bar{\Xi}^0}$ is consistent with the latest measurement result as mentioned in Sec.~\ref{intro}.
This tests the reliability of our estimations.
Lastly, in order to achieve a statistical sensitivity of $1 \% $, the data statistic must be greater than 100,000. Considering the influence of background level, detector efficiency, event reconstruction efficiency and other factors in each experiment, the actual required data sample size will be different and should be larger than we predicted.

\section{summary and outlook}
By studying the cascaded $J/\psi\to\Xi^0 \bar{\Xi}^0$, $\Xi^0\to\Lambda  \pi^0$, $\bar{\Xi}^0\to\bar{\Lambda} \gamma$ decays, the formulae of the angular distribution and the observed quantities of the polarization are arrived at, which can be used to measure the $\Xi $ decay asymmetric parameter in the future experiments. In particular, we estimate the statistical sensitivity of these parameters by considering the whole decay chain.
According to the estimation results, even a large asymmetric parameter value needs more than 100,000 data events to reach a the measurement accuracy at 1\%.

\section{acknowledgements}
The work is partly supported by the National Natural Science Foundation of China under Grants No. 12175244, No. 11875262, No. 11805037 and No. U1832121. National Key Research and Development Program of China under Contracts No. 2020YFA0406301.
\appendix
\section{Parameters}
The helicity amplitudes can be written in the form of a complex number as

\begin{eqnarray}\label{ComplexForm}
		A^J_{\lambda_1,\lambda_2}=a^J_{\lambda_1,\lambda_2}e^{i\xi_{\lambda_1,\lambda_2}},
\end{eqnarray}

1.	Expressions of real multipole parameters $Q^1_{i,j}$
\setlength\abovedisplayskip{3pt}
\setlength\belowdisplayskip{3pt}
\begin{eqnarray}\label{Q1expression}
	Q^{1}_{0,0} &=& a_{\frac{1}{2},\frac{1}{2}}^2 \sin ^2\theta _0+\frac{1}{4} a_{\frac{1}{2},-\frac{1}{2}}^2 (\cos 2 \theta _0+3),\nonumber\\
	Q^{1}_{0,0}Q^{1}_{0,2} &=& -\frac{a_{\frac{1}{2},-\frac{1}{2}} a_{\frac{1}{2},\frac{1}{2}} \sin 2 \theta _0\sin \Delta _a}{\sqrt{2}},\nonumber\\
	Q^{1}_{0,0}Q^{1}_{1,1} &=& \frac{1}{2} (2 a_{\frac{1}{2},\frac{1}{2}}^2+a_{\frac{1}{2},-\frac{1}{2}}^2)\sin ^2\theta _0,\nonumber\\
	Q^{1}_{0,0}Q^{1}_{1,3} &=& \frac{a_{\frac{1}{2},-\frac{1}{2}} a_{\frac{1}{2},\frac{1}{2}} \sin 2 \theta _0\cos \Delta _a}{\sqrt{2}},\nonumber\\
	Q^{1}_{0,0}Q^{1}_{2,0} &=&\frac{a_{\frac{1}{2},-\frac{1}{2}} a_{\frac{1}{2},\frac{1}{2}} \sin 2 \theta _0\sin \Delta _a}{\sqrt{2}},\nonumber\\
	Q^{1}_{0,0}Q^{1}_{2,2} &=& \frac{1}{2} (a_{\frac{1}{2},-\frac{1}{2}}^2-2 a_{\frac{1}{2},\frac{1}{2}}^2)\sin ^2\theta _0,\nonumber\\
	Q^{1}_{0,0}Q^{1}_{3,1} &=& -\frac{a_{\frac{1}{2},-\frac{1}{2}} a_{\frac{1}{2},\frac{1}{2}} \sin 2 \theta _0\cos \Delta _a}{\sqrt{2}},\nonumber\\
	Q^{1}_{0,0}Q^{1}_{3,3} &=&a_{\frac{1}{2},\frac{1}{2}}^2 \sin ^2\theta _0-\frac{1}{4} a_{\frac{1}{2},-\frac{1}{2}}^2 (\cos 2 \theta _0+3),
\end{eqnarray}
and others are equal to zero. \\

2.	The elements of $\rho^{\Lambda \bar{\Lambda}}$
\setlength\abovedisplayskip{3pt}
\setlength\belowdisplayskip{3pt}
\begin{eqnarray}\label{rhoLL}
		\rho^{\Lambda\bar{\Lambda}}_{\frac{1}{2}, \frac{1}{2}, \frac{1}{2},\frac{1}{2}}
		&=&Q^{1}_{0,0} (1+\alpha _{\Xi^0 }) (1-\alpha _{\bar{\Xi }^0})\nonumber\\
		&\times& \left\{\right.Q^{1}_{0,2} \sin \theta _2 \sin \phi _2+Q^{1}_{2,0} \sin \theta _1 \sin \phi _1\nonumber\\
		&+&Q^{1}_{2,2} \sin \theta _1 \sin \theta _2 \sin \phi _1 \sin \phi _2\nonumber\\
		&+&Q^{1}_{1,1} \sin\theta _1 \sin \theta _2 \cos \phi _1 \cos \phi _2\nonumber\\
		&+&Q^{1}_{1,3} \sin \theta _1 \cos \theta _2 \cos \phi _1\nonumber\\
		&+&Q^{1}_{3,1} \sin \theta _2 \cos \theta _1 \cos \phi _2\nonumber\\
		&+&Q^{1}_{3,3} \cos \theta _1 \cos \theta _2+1\left.\right\},\nonumber\\
		\rho^{\Lambda\bar{\Lambda}}_{\frac{1}{2}, \frac{1}{2}, -\frac{1}{2},\frac{1}{2}}
		&=&Q^{1}_{0,0} \sqrt{1-\alpha _{\Xi^0 }^2} e^{-i \Delta _b} (1-\alpha _{\bar{\Xi }^0})\nonumber\\
		&\times& \left\{\right.Q^{1}_{1,1} \sin \theta _2 \cos \phi _2 (\cos \theta _1 \cos \phi _1\nonumber\\
		&+&i \sin \phi _1)+Q^{1}_{1,3} \cos \theta _2 (\cos \theta _1 \cos \phi _1\nonumber\\
		&+&i \sin \phi _1)+Q^{1}_{2,0} \cos \theta _1 \sin \phi _1\nonumber\\
		&-&i Q^{1}_{2,2} \sin \theta _2 \sin \phi _2 \cos \phi _1\nonumber\\
		&+&Q^{1}_{2,2} \sin \theta _2 \cos \theta _1 \sin \phi _1 \sin \phi _2\nonumber\\
		&-&Q^{1}_{3,1} \sin \theta _1 \sin \theta _2 \cos \phi _2\nonumber\\
		&-&Q^{1}_{3,3} \sin \theta _1 \cos \theta _2-i Q^{1}_{2,0} \cos \phi _1\left.\right\},\nonumber\\
		\rho^{\Lambda\bar{\Lambda}}_{\frac{1}{2}, -\frac{1}{2}, \frac{1}{2},-\frac{1}{2}}
		&=&-Q^{1}_{0,0} (1+\alpha _{\Xi^0 }) (1+\alpha _{\bar{\Xi }^0}) \nonumber\\
		&\times&\left\{\right.-1+Q^{1}_{0,2} \sin \theta _2 \sin \phi _2 \nonumber\\&-&Q^{1}_{2,0} \sin \theta _1 \sin \phi _1\nonumber\\
		&+&Q^{1}_{2,2} \sin \theta _1 \sin \theta _2 \sin \phi _1 \sin \phi _2\nonumber\\
		&+&Q^{1}_{1,1} \sin \theta _1 \sin \theta _2 \cos \phi _1 \cos \phi _2\nonumber\\
		&+&Q^{1}_{1,3} \sin \theta _1 \cos \theta _2 \cos \phi _1\nonumber\\
		&+&Q^{1}_{3,1} \sin \theta _2 \cos \theta _1 \cos \phi _2\nonumber\\
		&+&Q^{1}_{3,3} \cos \theta _1 \cos \theta _2\left.\right\},\nonumber\\
		\rho^{\Lambda\bar{\Lambda}}_{\frac{1}{2}, -\frac{1}{2}, -\frac{1}{2},-\frac{1}{2}}
		&=&-Q^{1}_{0,0} \sqrt{1-\alpha _{\Xi^0 }^2} e^{-i \Delta _b} (\alpha _{\bar{\Xi }^0}+1) \nonumber\\
		&\times&\left\{\right.Q^{1}_{1,1} \sin \theta _2 \cos \phi _2 (\cos \theta _1 \cos \phi _1\nonumber\\
		&+&i \sin \phi _1)+Q^{1}_{1,3} \cos \theta _2 (\cos \theta _1 \cos \phi _1 \nonumber\\
		&+&i \sin \phi _1)-Q^{1}_{2,0} \cos \theta _1 \sin \phi _1\nonumber\\
		&-&i Q^{1}_{2,2} \sin \theta _2 \sin \phi _2 \cos \phi _1\nonumber\\
		&+&Q^{1}_{2,2} \sin \theta _2 \cos \theta _1 \sin \phi _1 \sin \phi _2\nonumber\\
		&-&Q^{1}_{3,1} \sin \theta _1 \sin \theta _2 \cos \phi _2\nonumber\\
		&-&Q^{1}_{3,3} \sin \theta _1 \cos \theta _2+i Q^{1}_{2,0} \cos \phi _1\left.\right\},\nonumber\\
		\rho^{\Lambda\bar{\Lambda}}_{-\frac{1}{2}, \frac{1}{2}, \frac{1}{2},\frac{1}{2}}
		&=&Q^{1}_{0,0} \sqrt{1-\alpha _{\Xi^0}^2} e^{i \Delta _b} (1-\alpha _{\bar{\Xi }^0})\nonumber\\
		&\times& \left\{\right.Q^{1}_{1,1} \sin \theta _2 \cos \phi _2 (\cos \theta _1 \cos \phi _1\nonumber\\
		&-&i \sin \phi _1)+Q^{1}_{1,3} \cos \theta _2 (\cos \theta _1 \cos \phi _1 \nonumber\\
		&-&i \sin \phi _1)+Q^{1}_{2,0} \cos \theta _1 \sin \phi _1\nonumber\\
		&+&i Q^{1}_{2,2} \sin \theta _2 \sin \phi _2 \cos \phi _1\nonumber\\
		&+&Q^{1}_{2,2} \sin \theta _2 \cos \theta _1 \sin \phi _1 \sin \phi _2\nonumber\\
		&-&Q^{1}_{3,1} \sin \theta _1 \sin \theta _2 \cos \phi _2\nonumber\\
		&-&Q^{1}_{3,3} \sin \theta _1 \cos \theta _2+i Q^{1}_{2,0} \cos \phi _1\left.\right\},\nonumber\\
		\rho^{\Lambda\bar{\Lambda}}_{-\frac{1}{2}, \frac{1}{2}, -\frac{1}{2},\frac{1}{2}}
		&=&Q^{1}_{0,0} (1-\alpha _{\Xi^0 }) (1-\alpha _{\bar{\Xi }^0})\nonumber\\
		&\times& \left\{\right.1+Q^{1}_{0,2} \sin \theta _2 \sin \phi _2\nonumber\\
		&-&Q^{1}_{2,0} \sin \theta _1 \sin \phi _1\nonumber\\
		&-&Q^{1}_{2,2} \sin \theta _1 \sin \theta _2 \sin \phi _1 \sin \phi _2\nonumber\\
		&-&Q^{1}_{1,1} \sin \theta _1 \sin \theta _2 \cos \phi _1 \cos \phi _2\nonumber\\
		&-&Q^{1}_{1,3} \sin \theta _1 \cos \theta _2 \cos \phi _1\nonumber\\
		&-&Q^{1}_{3,1} \sin \theta _2 \cos \theta _1 \cos \phi _2\nonumber\\
		&-&Q^{1}_{3,3} \cos \theta _1 \cos \theta _2\left.\right\},\nonumber\\
		\rho^{\Lambda\bar{\Lambda}}_{-\frac{1}{2}, -\frac{1}{2}, \frac{1}{2},-\frac{1}{2}}
		&=&-Q^{1}_{0,0} \sqrt{1-\alpha _{\Xi^0 }^2} e^{i \Delta _b} (\alpha _{\bar{\Xi }^0}+1)\nonumber\\
		&\times& \left\{\right.Q^{1}_{1,1} \sin \theta _2 \cos \phi _2 (\cos \theta _1 \cos \phi _1\nonumber\\
		&-&i \sin \phi _1)+Q^{1}_{1,3} \cos \theta _2 (\cos \theta _1 \cos \phi _1\nonumber\\
		&-&i \sin \phi _1)-Q^{1}_{2,0} \cos \theta _1 \sin \phi _1\nonumber\\
		&+&i Q^{1}_{2,2} \sin \theta _2 \sin \phi _2 \cos \phi _1\nonumber\\
		&+&Q^{1}_{2,2} \sin \theta _2 \cos \theta _1 \sin \phi _1 \sin \phi _2\nonumber\\
		&-&Q^{1}_{3,1} \sin \theta _1 \sin \theta _2 \cos \phi _2\nonumber\\
		&-&Q^{1}_{3,3} \sin \theta _1 \cos \theta _2-i Q^{1}_{2,0} \cos \phi _1\left.\right\},\nonumber\\
		\rho^{\Lambda\bar{\Lambda}}_{-\frac{1}{2}, -\frac{1}{2}, -\frac{1}{2},-\frac{1}{2}}
		&=&Q^{1}_{0,0} (1-\alpha _{\Xi^0 })(1+\alpha _{\bar{\Xi }^0})\nonumber\\
		&\times& \left\{\right.1-Q^{1}_{0,2} \sin \theta _2 \sin \phi _2\nonumber\\
		&-&Q^{1}_{2,0} \sin \theta _1 \sin \phi _1\nonumber\\
		&+&Q^{1}_{2,2} \sin \theta _1 \sin \theta _2 \sin \phi _1 \sin \phi _2\nonumber\\
		&+&Q^{1}_{1,1} \sin \theta _1 \sin \theta _2 \cos \phi _1 \cos \phi _2\nonumber\\
		&+&Q^{1}_{1,3} \sin \theta _1 \cos \theta _2 \cos \phi _1\nonumber\\
		&+&Q^{1}_{3,1} \sin \theta _2 \cos \theta _1 \cos \phi _2\nonumber\\
		&+&Q^{1}_{3,3} \cos \theta _1 \cos \theta _2\left.\right\},\nonumber\\
\end{eqnarray}
and the unlisted are equal to zero.\\

3.	Expressions of real multipole parameters $Q^2_{i,j}$
\begin{eqnarray}\label{Q2expression}
		Q^{2}_{0,0}
		&=&\frac{1}{4} Q^{1}_{0,0} \left\{\right.1+Q^{1}_{2,0} \alpha _{\Xi^0 } \sin \theta _1 \sin \phi _1\nonumber\\
		&+&\alpha _{\bar{\Xi }^0} [-\alpha _{\Xi^0 } (Q^{1}_{1,1} \sin \theta _1 \sin \theta _2 \cos \phi _1 \cos \phi _2\nonumber\\
		&+&Q^{1}_{1,3} \sin \theta _1 \cos \theta _2 \cos \phi _1 \nonumber\\
		&+&\sin \theta _2 (Q^{1}_{2,2} \sin \theta _1 \sin \phi _1 \sin \phi _2\nonumber\\
		&+&Q^{1}_{3,1} \cos \theta _1 \cos \phi _2)+Q^{1}_{3,3} \cos \theta _1 \cos \theta _2)\nonumber\\
		&-&Q^{1}_{0,2} \sin \theta _2 \sin \phi _2]\left.\right\},\nonumber\\
		Q^{2}_{0,0}Q^{2}_{0,3}
		&=&\frac{1}{4} Q_{0,0} \left\{\right.-\alpha _{\bar{\Xi }^0} (Q_{2,0} \alpha _{\Xi ^0} \sin \theta _1 \sin \phi _1+1)\nonumber\\
		&+&\alpha _{\Xi^0} [Q_{2,2} \sin \theta _1 \sin \theta _2 \sin \phi _1 \sin \phi _2\nonumber\\
		&+&Q_{1,1} \sin \theta _1 \sin \theta _2 \cos \phi _1 \cos \phi _2\nonumber\\
		&+&Q_{1,3} \sin \theta _1 \cos \theta _2 \cos \phi _1\nonumber\\
		&+&Q_{3,1} \sin \theta _2 \cos \theta _1 \cos \phi _2+Q_{3,3} \cos \theta _1 \cos \theta _2]\nonumber\\
		&+&Q_{0,2} \sin \theta _2 \sin \phi _2\left.\right\},\nonumber\\
		Q^{2}_{0,0}Q^{2}_{1,0}
		&=&-\frac{1}{4} Q_{0,0} \sqrt{1-\alpha _{\Xi^0 }^2} \left\{\right.\alpha _{\bar{\Xi }^0} [Q_{1,1} \sin \theta _2 \cos \phi _2\nonumber\\
		&\times& (\cos \theta _1 \cos \phi _1 \cos \Delta _b+\sin \phi _1 \sin \Delta _b)\nonumber\\
		&+&Q_{1,3} \cos \theta _2 (\cos \theta _1 \cos \phi _1 \cos \Delta _b\nonumber\\
		&+&\sin \phi _1 \sin \Delta _b)+\sin \theta _2 (Q_{2,2} \sin \phi _2\nonumber\\
		&\times& (\cos \theta _1 \sin \phi _1 \cos \Delta _b-\cos \phi _1 \sin \Delta _b)\nonumber\\
		&-&Q_{3,1} \sin \theta _1 \cos \phi _2 \cos \Delta _b)\nonumber\\
		&-&Q_{3,3} \sin \theta _1 \cos \theta _2 \cos \Delta _b]\nonumber\\
		&+&Q_{2,0} (\cos \phi _1 \sin \Delta _b-\cos \theta _1 \sin \phi _1 \cos \Delta _b)\left.\right\},\nonumber\\
		Q^{2}_{0,0}Q^{2}_{1,3}
		&=&\frac{1}{4} Q^{1}_{0,0} \sqrt{1-\alpha _{\Xi^0 }^2} [-Q^{1}_{2,0} \cos \theta _1 \sin \phi _1 \alpha _{\bar{\Xi }^0} \nonumber\\
		&\times&\cos \Delta _b+Q^{1}_{1,1} \sin \theta _2 \cos \phi _2 \nonumber\\
		&\times&(\cos \theta _1 \cos \phi _1 \cos \Delta _b+\sin \phi _1 \sin \Delta _b)\nonumber\\
		&+&Q^{1}_{1,3} \cos \theta _2 (\cos \theta _1 \cos \phi _1 \cos \Delta _b\nonumber\\
		&+&\sin \phi _1 \sin \Delta _b)\nonumber\\
		&-&Q^{1}_{2,2} \sin \theta _2 \sin \phi _2 \cos \phi _1 \sin \Delta _b\nonumber\\
		&+&Q^{1}_{2,2} \sin \theta _2 \cos \theta _1 \sin \phi _1 \sin \phi _2 \cos \Delta _b\nonumber\\
		&-&Q^{1}_{3,1} \sin \theta _1 \sin \theta _2 \cos \phi _2 \cos \Delta _b\nonumber\\
		&+&Q^{1}_{2,0} \cos \phi _1 \alpha _{\bar{\Xi }^0} \sin \Delta _b\nonumber\\
		&-&Q^{1}_{3,3} \sin \theta _1 \cos \theta _2 \cos \Delta _b],\nonumber\\
		Q^{2}_{0,0}Q^{2}_{2,0}
		&=&\frac{1}{4} Q^{1}_{0,0} \sqrt{1-\alpha _{\Xi ^0}^2} \left\{\right.\alpha _{\bar{\Xi }^0} [Q^{1}_{1,1} \sin \theta _2 \cos \phi _2 \nonumber\\
		&\times&(\sin \phi _1 \cos \Delta _b-\cos \theta _1 \cos \phi _1 \sin \Delta _b)\nonumber\\
		&+&Q^{1}_{1,3} \cos \theta _2 (\sin \phi _1 \cos \Delta _b\nonumber\\
		&-&\cos \theta _1 \cos \phi _1 \sin \Delta _b)+\sin \theta _2 (-Q^{1}_{2,2} \sin \phi _2\nonumber\\
		&\times& (\cos \theta _1 \sin \phi _1 \sin \Delta _b+\cos \phi _1 \cos \Delta _b)\nonumber\\
		&+&Q^{1}_{3,1} \sin \theta _1 \cos \phi _2 \sin \Delta _b)\nonumber\\
		&+&Q^{1}_{3,3} \sin \theta _1 \cos \theta _2 \sin \Delta _b]\nonumber\\
		&+&Q^{1}_{2,0} (\cos \theta _1 \sin \phi _1 \sin \Delta _b+\cos \phi _1 \cos \Delta _b)\left.\right\},\nonumber\\
		Q^{2}_{0,0}Q^{2}_{2,3}
		&=&\frac{1}{4} Q^{1}_{0,0} \sqrt{1-\alpha _{\Xi ^0}^2} [-Q^{1}_{2,0} \cos \theta _1 \sin \phi _1 \alpha _{\bar{\Xi }^0} \nonumber\\
		&\times&\sin \Delta _b-Q^{1}_{2,0} \cos \phi _1 \alpha _{\bar{\Xi }^0} \cos \Delta _b\nonumber\\
		&+&Q^{1}_{1,1} \sin \theta _2 \cos \phi _2 (\cos \theta _1 \cos \phi _1 \sin \Delta _b\nonumber\\
		&-&\sin \phi _1 \cos \Delta _b)+Q^{1}_{1,3} \cos \theta _2 (\cos \theta _1 \cos \phi _1\nonumber\\
		&\times& \sin \Delta _b-\sin \phi _1 \cos \Delta _b)+Q^{1}_{2,2} \sin \theta _2 \nonumber\\
		&\times&\sin \phi _2 \cos \phi _1 \cos \Delta _b+Q^{1}_{2,2} \sin \theta _2 \cos \theta _1 \nonumber\\
		&\times&\sin \phi _1 \sin \phi _2 \sin \Delta _b-Q^{1}_{3,1} \sin \theta _1 \sin \theta _2\nonumber\\
		&\times& \cos \phi _2 \sin \Delta _b-Q^{1}_{3,3} \sin \theta _1 \cos \theta _2 \sin \Delta _b],\nonumber\\
		Q^{2}_{0,0}Q^{2}_{3,0}
		&=&\frac{1}{4} Q^{1}_{0,0} \left\{\right.\alpha _{\Xi ^0} (1-Q^{1}_{0,2} \sin \theta _2 \sin \phi _2 \alpha _{\bar{\Xi }^0})\nonumber\\
		&-&\alpha _{\bar{\Xi }^0} [Q^{1}_{1,1} \sin \theta _1 \sin \theta _2 \cos \phi _1 \cos \phi _2\nonumber\\
		&+&Q^{1}_{1,3} \sin \theta _1 \cos \theta _2 \cos \phi _1+\sin \theta _2 (Q^{1}_{2,2} \sin \theta _1\nonumber\\
		&\times& \sin \phi _1 \sin \phi _2+Q^{1}_{3,1} \cos \theta _1 \cos \phi _2)\nonumber\\
		&+&Q^{1}_{3,3} \cos \theta _1 \cos \theta _2]+Q^{1}_{2,0} \sin \theta _1 \sin \phi _1\left.\right\},\nonumber\\
		Q^{2}_{0,0}Q^{2}_{3,3}
		&=&\frac{1}{4} Q^{1}_{0,0} [-\alpha _{\Xi^0 } (\alpha _{\bar{\Xi }^0}-Q^{1}_{0,2} \sin \theta _2 \sin \phi _2)\nonumber\\
		&-&Q^{1}_{2,0} \sin \theta _1 \sin \phi _1 \alpha _{\bar{\Xi }^0}+Q^{1}_{2,2} \sin \theta _1 \sin \theta _2\nonumber\\
		&\times& \sin \phi _1 \sin \phi _2+Q^{1}_{1,1} \sin \theta _1 \sin \theta _2 \cos \phi _1 \cos \phi _2\nonumber\\
		&+&Q^{1}_{1,3} \sin \theta _1 \cos \theta _2 \cos \phi _1+Q^{1}_{3,1} \sin \theta _2 \cos \theta _1\nonumber\\
		&\times& \cos \phi _2+Q^{1}_{3,3} \cos \theta _1 \cos \theta _2],
\end{eqnarray}
and the others are equal to zero.

\end{document}